\documentclass[fleqn,12pt,twoside]{article}
\usepackage{espcrc1}

\usepackage{graphicx}

\newcommand{\AmS}{{\protect\the\textfont2
  A\kern-.1667em\lower.5ex\hbox{M}\kern-.125emS}}

\title{Correlations and Fluctuations, A Summary of Quark Matter 2002}

\author{Scott Pratt\address{Department of Physics and Astronomy, Michigan State
        University,\\ East Lansing, MI 48824, USA} \thanks{This work was
        supported by the United States National Science Foundation, Grant No.
        PHY-00-70818.}}
       
\begin{document}

\maketitle

\begin{abstract}
Results for correlations and fluctuations presented at Quark Matter 2002 are
summarized. These results include Hanbury-Brown Twiss interferometry of a wide
variety of species, large scale fluctuations and correlations in $p_t$ and
multiplicity, and charge fluctuations and charge balance functions.
\end{abstract}

\section{Introduction}

A phenomenal number of new results have been presented at this meeting
regarding fluctuations and correlations. With the analysis of SPS results for
$\pi-\pi$ Hanbury-Brown Twiss (HBT) correlations as a function of energy, and
with the maturing of the RHIC results, a consistent picture has developed of
the evolving dynamics from AGS energies to RHIC. Additionally, several new
probes have been analyzed and have been presented at this meeting, including
$\gamma \gamma$ correlations. Fluctuations in $p_t$, multiplicity and isospin,
derived from both the SPS and RHIC, hint at novel behaviors. Finally, charge
correlations measured at RHIC suggest a delayed production of charge, and
therefore a delayed hadronization.  After being taken by surprise, the theory
community has made significant progress toward understanding these measurements
and reducing the list of possible interpretations.

Results on small relative momentum correlations are reviewed in the next
section, with an emphasis on describing the HBT puzzle. The subsequent section
reviews $p_t$ and multiplicity fluctuations, while isospin fluctuations, charge
fluctuations and charge balance functions are reviewed in
Sec. \ref{sec:chargeflucbalance}.

\section{Two-Particle Correlations at Small Relative Momentum}

\begin{figure}
\centerline{\includegraphics[width=0.84\textwidth]{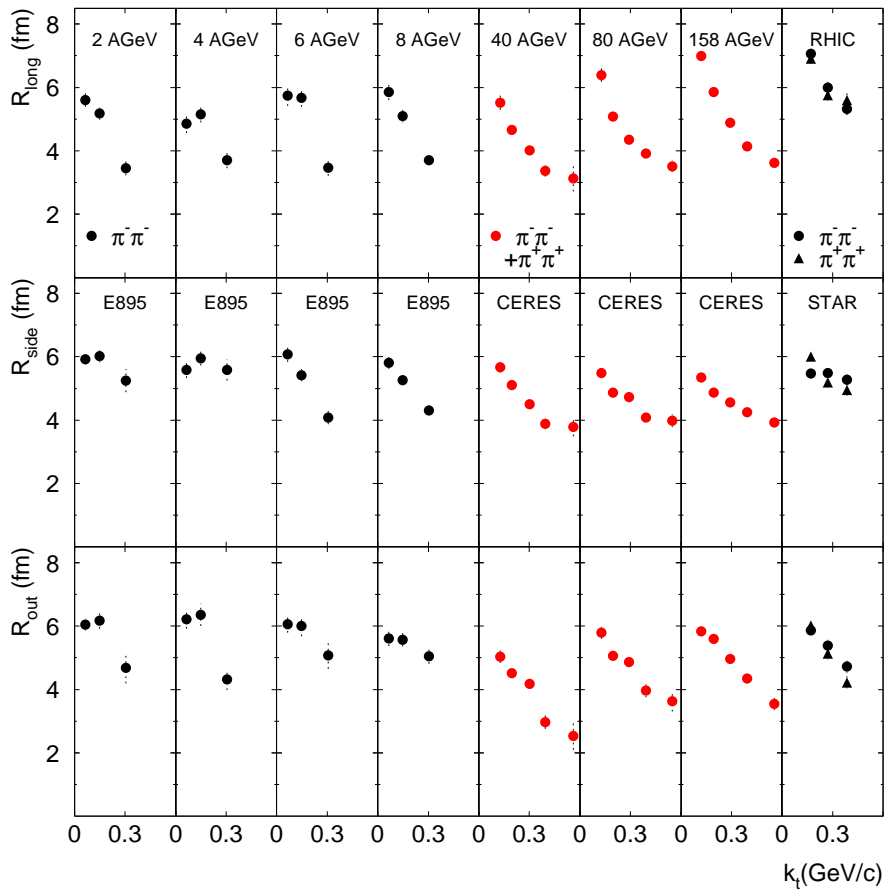}}
\caption{\label{fig:ceressummary}The three dimensions of the source are
  illustrated as a function of transverse momentum for several beam
  energies. Only $R_{\rm long}$ changes significantly for higher beam energies.
}
\end{figure}

A principal motivation for HBT analyses was the hope to view a long-lived
source resulting from the low pressure associated with a first-order phase
transition to the the quark-gluon plasma (QGP)
\cite{prattprd86,bertschdrop,rischkehbt}. The CERES collaboration presented
results, shown in Fig.  \ref{fig:ceressummary}, of both their results at the
SPS along with AGS and RHIC results \cite{qm2002tilsner,ceresadamova}.  A
long-lived source would have emitted pions continuously over a time of the
order of 10-20 fm/c, which would result in the outward size $R_{\rm out}$
being larger than $R_{\rm side}$ by an amount of the order of 10 fm. Figure
\ref{fig:ceressummary} shows that such a long-lived source was not produced at
any energy. In fact, the outward and sidewards dimensions show little
dependence with beam energy. Consistent measurements of the three source
dimensions at RHIC were reported in this meeting by STAR, PHENIX and PHOBOS,
with the ratio of $R_{\rm out}/R_{\rm side}$ varying from $\approx 0.85$ for
PHENIX \cite{qm2002enokizono} to $\approx 1.15$ for PHOBOS \cite{qm2002manly}.
Because of the inherent ambiguity between the temporal and spatial dimensions
in $R_{\rm out}$, one can only place an upper limit on the duration of the
emission of $\Delta\tau<10$ fm/c. Given the large transverse size, it appears
that the breakup hypersurface is extremely space-like, i.e. emissions from the
edge and the center occur within a few fm/c of one another.

The values reported for $R_{\rm long}$ are as remarkable as the $R_{\rm
out}/R_{\rm side}$ measurements. Whereas $R_{\rm out}/R_{\rm side}$ provides
insight into the duration of emission, $R_{\rm long}$ gives a measure of the
emission time.  HBT measures the spatial extent from which pions of a given
rapidity might be emitted. Assuming that the emitting sources are well spread
out in rapidity, the size is determined by the temperature and the velocity
gradient, $R_{\rm long}\sim v_{\rm th}/(dv/dz)$, where $v_{\rm th}$ is the
thermal velocity and $dv/dz$ is the velocity gradient. In a Bjorken expansion,
the matter does not accelerate along the $z$ axis which means that the velocity
gradient is determined by $\tau$, $dv/dz=1/\tau$. Although the values for
$R_{\rm long}$ reported by STAR, PHENIX and PHOBOS are very large compared to
hadronic length scales, they still fall short of expectations, and suggest
emission times of 10 fm/c, rather than the expected 15-20 fm/c. Unless the
breakup temperatures are anomalously cold, or unless the matter accelerated in
the longitudinal direction, explaining this result requires an extremely rapid
transverse acceleration. Blast wave analyses \cite{qm2002retiere} of $\pi-\pi$
HBT reveal sources which have expanded to an outer radius of 13 fm within this
10 fm/c, and with an outer velocity of $\sim 0.7c$ as determined by comparing
pion and proton spectra. Given that the original radius is near 6 fm, this
implies an extremely rapid acceleration.

\begin{figure}
\centerline{\includegraphics[angle=90,width=0.9\textwidth]{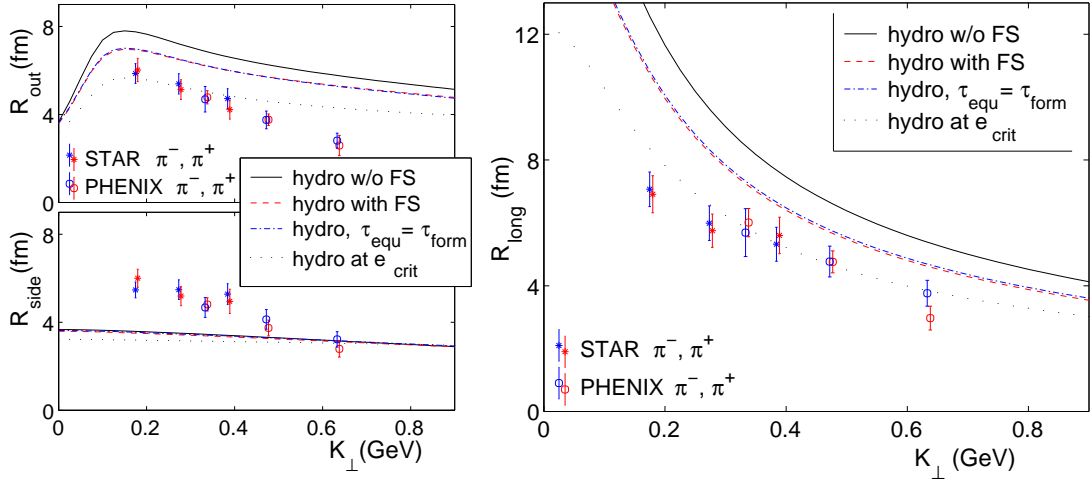}}
\caption{\label{fig:kolbhbt}Theoretical predictions of Kolb et al. compared
  to STAR results. As indicative of several other treatments, the inclusion of
  the final hadronic rescattering pushes the theoretical predictions further
  from the results.}
\end{figure}

The inability of numerical models of the dynamics to account for both the small
values of the mean emission time, $\langle \tau\rangle$, and the duration of
the emission, $\Delta \tau$, is referred to as the HBT puzzle. Several attempts
have been made at reproducing the RHIC results using a variety of hydrodynamic,
Boltzmann and hybrid treatments. All consistently overestimate $R_{\rm long}$
and $R_{\rm out}/R_{\rm side}$
\cite{teaneyhydro,qm2002huovinen,qm2002soff,qm2002kolb}. Figure
\ref{fig:kolbhbt} displays results from Kolb which come somewhat closer to the
result than the other reference work cited above. It should be emphasized that
the HBT puzzle is not particularly new to RHIC. As HBT statistics have
increased form the SPS, it is clear that the emission duration was moderately
over-predicted by RQMD at SPS energies, with measurements of $R_{\rm
out}/R_{\rm side}$ varying from near unity for CERES \cite{qm2002tilsner} to
$\sim 1.15$ for NA49 \cite{qm2002blume}.

Two-kaon correlations from PHENIX and from NA49 were also reported at this
meeting \cite{qm2002blume,qm2002enokizono}. It appears that the source size for
kaons is somewhat larger than predicted by extrapolating from the pion source
sizes assuming blast wave kinematics. However, it should be pointed out that
the NA49 measurement is somewhat larger than that previously reported by NA44
\cite{na44kk}, and that the PHENIX measurement is preliminary.

Aside from the kaon measurements, the overall behavior of HBT measurements
appears to follow the systematics of a blast wave, i.e. a sudden dissolution at
a low temperature, $\sim 110 MeV$, after a rapid expansion to a high velocity,
$\sim 0.7c$. In addition to the $R_{\rm out}/R_{\rm side}$ ratio and the small
value of $R_{\rm long}$, the blast wave picture of sudden explosion and
emission explains the $p_t$ dependence of the source sizes
\cite{qm2002enokizono,qm2002blume,qm2002tilsner,qm2002ray}, and may also
explain the results of STAR where two-pion HBT was performed relative to the
reaction plane \cite{qm2002lopeznoriega}. There, the source appears to be
out-of-plane extended, i.e. the source appears to retain the shape of the
original almond resulting from overlapping spheres at non-zero impact
parameter. If the emission were long-lived, the differential expansion would
have overcompensated for the initial deformation and resulted in an
in-plane-extended shape.

One consequence of the smaller-than-expected values of $R_{\rm long}$ and
$R_{\rm out}$ is that the phase space density must have been rather high at
breakup. In fact, by combining the spectra and HBT measurements, one can
extract the phase space density in a model-independent way
\cite{bertschphasespace}. This analysis was performed for STAR results
\cite{qm2002ray}, and the resulting phase space density suggests that pionic
phase space is overpopulated by nearly a factor of two at low momentum as
compared to a thermally equilibrated gas where the pion density is determined
solely by the temperature. The phase space density increases markedly with
centrality and is significantly higher than was extracted for central
collisions at SPS energies \cite{na49phasespace}.  It should be emphasized that
the reported phase space values are averaged over coordinate space,
\begin{equation}
\bar{f}({\bf p}) \equiv \frac{\int d^3x f({\bf x},{\bf p})f({\bf x},{\bf p})}
    {\int d^3x f({\bf x},{\bf p})},
\end{equation}
For a Gaussian source the peak phase space density is larger by $2\sqrt{2}$ at
$x=0$ than the average derived from the sum rule. This suggests that the break
up density is a significant fraction of what is required for novel Bose
coherence effects.

\begin{figure}
\centerline{\includegraphics[width=0.8\textwidth]{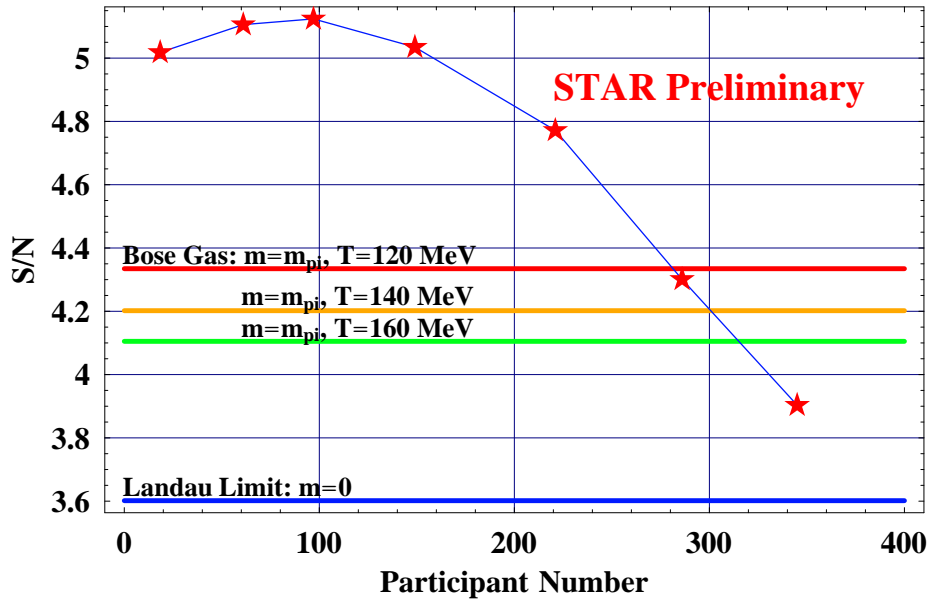}}
\caption{\label{fig:cramer_entropy} The entropy per pion as extracted from HBT
  and spectra. For the most central collisions, the entropy per pion appears
  low which puts doubt into assumptions about early equilibration.}
\end{figure}

A high phase-space density necessitates a lower entropy. In a nearly
model-independent fashion, the entropy carried by the pions was extracted from
STAR results \cite{qm2002cramerposter}. The entropy per pion appears to be in
the neighborhood of three units which is significantly lower than that expected
for a Bose gas as can be seen in Fig. \ref{fig:cramer_entropy}. As hydrodynamic
treatments over-predict the source sizes, while fitting the spectra
\cite{teaneyhydro}, they thus over-predict the entropy. Since hydrodynamic
treatments produce minimal entropy in their evolution, one must conclude that
the initial entropy of the system in the hydrodynamic treatments was
overestimated. This suggests that the initial state was not that of an
equilibrated quark-gluon plasma, but might have instead involved energy in
coherent forms, e.g. classical gluon fields, or may have excited fewer degrees
of freedom, e.g. under-populating quark degrees of freedom.

The correlation functions presented at this conference for non-identical
particles were shockingly superior to what had been previously
measured. Following the ideas of Lednicky \cite{lednicky}, correlations were
compared for $p_{a,{\rm out}}-p_{b,{\rm out}}>0$ to those with $p_{a,{\rm
out}}-p_{b,{\rm out}}<0$. Whereas the two correlations must be identical when
the particles are indistinguishable, the difference offers insight into the
relative positions of the emitted particles when the two particles are
different. STAR presented results for $\pi-K$ and $\pi-p$ pairs
\cite{qm2002retiere} which showed that the heavier particles were emitted
ahead of the lighter particles consistent with a blast wave interpretation.
Results from NA49 for $\pi-p$ pairs demonstrated the same behavior
\cite{qm2002blume}.

In this meeting correlations were reported for several new pairs which had not
previously been measured in heavy ion collisions. NA49 displayed results for
both $\Lambda\Lambda$ and $\Lambda p$ correlations \cite{qm2002blume}. The
$\Lambda\Lambda$ results showed a flat correlation, which means that they are
not useful for source-size determination, but can put limits on the
$\Lambda\Lambda$ scattering length. This is important as it sheds light on
whether there is a low-energy resonance associated with the $H_0$
di-baryon. The $\Lambda-p$ correlation function exhibited a peak near zero
momentum which was consistent with expectations for a three to four Fermi
source, though the size was difficult to ascertain without knowing the
$\lambda$ parameter which depends on the fraction of $\Lambda$s from decays of
heavier hyperons. With greater statistics, this measurement offers the
possibility of determining whether $\Lambda$s are emitted ahead or behind the
protons.

The Coulomb interaction between the residual source and the pairs used for HBT
has always clouded the interpretation of experiments. However, $K_s-K_s$
interferometry sidesteps the issue since the $K_s$ is neutral. A first
measurement of the $K_s-K_s$ correlation function was presented by STAR
\cite{qm2002lopeznoriega}. Although the statistics were insufficient for a
detailed analysis, this preliminary result was extremely encouraging.

\begin{figure}
\centerline{\includegraphics[width=0.6\textwidth]{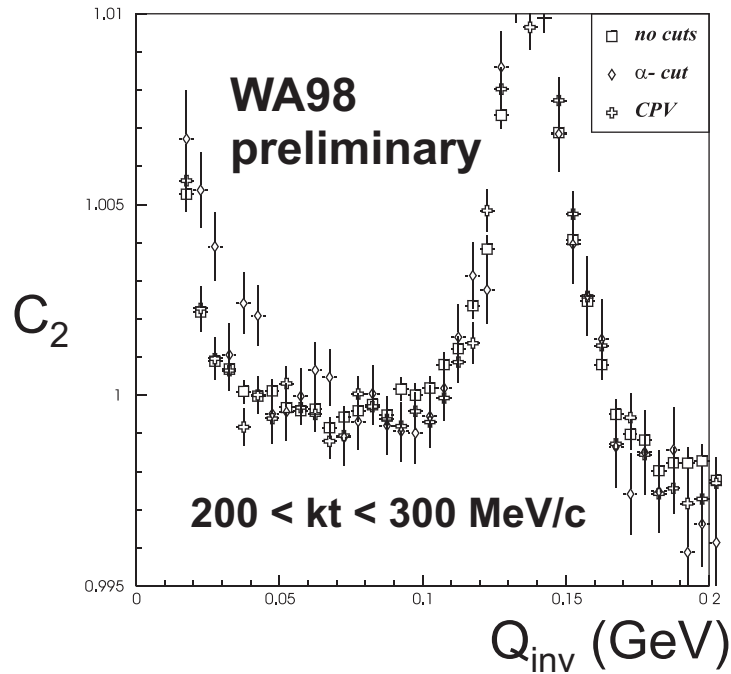}}
\caption{\label{fig:wa982gamma}The two-photon correlation function from WA98 is
shown with and without cuts for cluster splitting in the calorimeter.}
\end{figure}

The most surprising result of the conference was the presentation of two-photon
HBT by WA98 \cite{qm2002mohanty,wa98aggarwal}. The height of the correlation
function shown in Fig. \ref{fig:wa982gamma} has a height which requires that
10\% of the photons must be direct photons, i.e. not from $\pi_0$ decay or
dalitz decays from long-lived particles such as the $\eta$. In fact, by
measuring the inclusive photon spectrum, one can use the intercept of the
correlation function, $1+\lambda$, to infer the direct photon spectrum,
\begin{equation}
\frac{dN_{\rm direct}}{d^3p}=\sqrt{\lambda^{\gamma\gamma}(p_t)}\frac{dN_{\rm
inclusive}}{d^3p}.
\end{equation}
This result was surprising because the $\lambda$ parameter was expected to be
much smaller. If the fraction of direct photons to inclusive photons were 1\%
rather than 10\% the height of the correlation function would have been smaller
by a factor of 100, and the statistical requirement for events would have
increased by a factor of $10^4$. 

I conclude this section with a brief summary of theoretical attempts at solving
the HBT puzzle. Most importantly, it should be emphasized that there is
consensus that hydrodynamics with a Boltzmann or cascade afterburner will
over-predict source sizes
\cite{teaneyhydro,qm2002soff,qm2002kolb,qm2002huovinen,qm2002hirano,hiranohydro}. However,
it is not yet clear that fitting the data would not be possible if an extreme
equation of state were used, e.g. one with a speed of sound greater than
$1/\sqrt{3}$. Several careful studies have been performed to ensure that
experimental resolution and Coulomb corrections have been properly treated
\cite{qm2002tilsner,qm2002heffnerposter,qm2002brownposter}. The effects of
viscosity have just begun to be included, and it seems that they will provide
only a modest correction in order to be consistent with elliptic flow results
which preclude large viscosities \cite{qm2002teaney,qm2002heinz}. The effect of
modifying chemical rates and cross sections has also been investigated with
only modest changes in HBT results \cite{qm2002hirano,qm2002soff}. Finally, the
efforts at comparing theory to experiment have become more meaningful as
theoretical models have generated correlation functions which are then fit for
Gaussian parameters, rather than comparing radii generated from looking at
source functions. It appears that the HBT puzzle indeed derives from comparing
``apples to apples''.

In addition to listing all the investigations presented in this conference, it
is worthwhile to list the issues not yet addressed which might shed light on
the HBT puzzle. First, it should be pointed out that the hydrodynamic
treatments all assume a Bjorken space-time geometry, i.e. the matter coasts
along the beam axis. If the matter accelerates significantly due to the finite
size in the beam direction, the extrapolations for $\langle\tau\rangle$ at
breakup might be too small. This issue could be settled by three-dimensional
hydrodynamic studies. Secondly, three-body interactions such as the residual
Coulomb interaction have not been totally ruled out as being unimportant,
although schematic calculations suggest they are indeed unimportant, and the
equivalence of $\pi^+\pi^+$ and $\pi^-\pi^-$ results also points to their
insignificance. The most interesting possible explanation of the $R_{\rm
out}/R_{\rm side}$ puzzle might lay in a reduced emissivity of the surface,
which would allow the matter to fall apart suddenly. The emissivity might be
reduced in the case of a super-cooled phase. The problem of how quarks might
escape microscopically from a QGP region and form mesons has been considered
from the point of fissioning flux tubes
\cite{bannerjeeglendenningmatsui}. Effects of an increased emissivity were
reported on at this conference \cite{qm2002padula}. As a final possibility, it
may be that the picture of the dynamics assumed in cascade or Boltzmann
algorithms is fundamentally flawed due to the neglect of quantum
effects. Again, this seems unlikely, but it should not be forgotten that the
thermal wavelengths at breakup correspond to volumes with several particles. In
order to explain the HBT puzzle, these issues must somehow account for
corrections of the order of 30\%.

\section{Multiplicity correlations and $p_t$ fluctuations}
\begin{figure}
\centerline{\includegraphics[angle=270,width=0.60\textwidth]{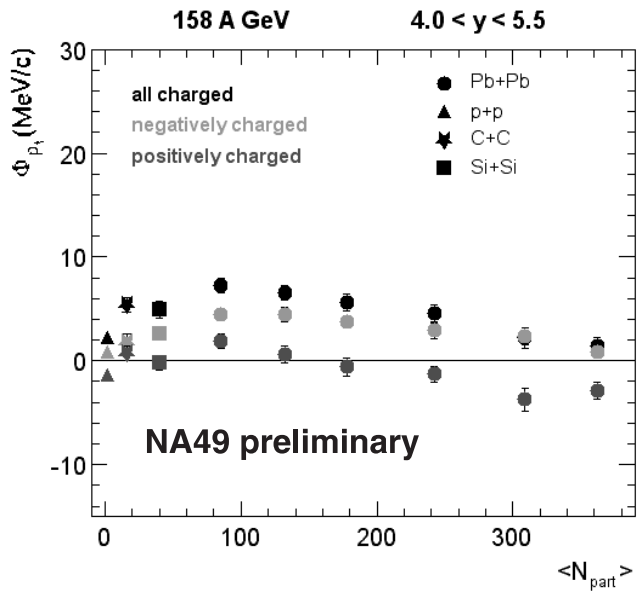}}
\caption{\label{fig:na49_ptfluc} Fluctuations in
  $p_t$ from the SPS are shown as a function of centrality for all charges or
  for same-sign particles.}
\end{figure}

Fluctuations in multiplicity or transverse momentum have been proposed as
signatures for phase separation \cite{seibert,wieand} or critical fluctuations
\cite{stephanov}. Of course, such fluctuations are also affected by jets and
jet quenching. In this conference we have seen new results for $p_t$
fluctuations from NA49 \cite{qm2002blume}, PHENIX \cite{qm2002mitchell} and
STAR \cite{qm2002ray}. The results from NA49, shown in
Fig. \ref{fig:na49_ptfluc}, illustrate that the fluctuation is smaller for
same-sign particles. This may be due to the positive correlation between
opposite sign particles from resonance decays. HBT effects, jets and the 
distorted shapes of the nuclei are other potential correlations which would
give $p_t$ fluctuations. 

The value of $\Phi_{p_t}$ increases if there is an increased correlation in
$p_t$ between given particles or if the number of particles with which one is
correlated increases. By measuring the kinematic range of the correlation, one
should gain insight into the source of the non-zero $\Phi_{p_t}$. The
fluctuations measured by PHENIX were even smaller than those reported by NA49,
while the fluctuations measured by STAR were significantly larger. There seems
to be a trend that the fluctuations are larger for larger acceptances, which
suggests that the correlation may span more than a unit of rapidity.

A wavelet analysis, which can measure the same class of correlations as $p_t$
fluctuations, was presented by the STAR collaboration in this meeting
\cite{qm2002kopytineposter}. For low multiplicities, the correlations were
consistent with HIJING with jets, while at high multiplicities, the results were
smaller and were consistent with HIJING without jets. This emphasizes the point
that jet quenching should be manifest in low $p_t$ observables. Perhaps,
experiment could decide not only the existence but the nature of jet quenching,
i.e. determining whether the jet energy is transformed into more particles
through radiation or whether it is shared among the collision partners.

\section{Isospin fluctuations and correlations arising from charge
  conservation}
\label{sec:chargeflucbalance}

Large fluctuations in the fraction of neutral pions to total pions has been
suggested as a signal for coherent emission of pions which might result from a
disoriented chiral condensate \cite{rajagopalwilczek}. Unfortunately, this is a
tremendously difficult measurement due to the fact that neutral pions decay
into two photons, making the exact counting of soft pions extremely difficult,
especially in a central collision. WA98 presented an analysis showing no
evidence for large fluctuations \cite{qm2002mohanty}. However, one should not
jump to the conclusion that the observations were entirely consistent with
random emissions until a more detailed theoretical investigation is
performed. In this conference, new ideas were also presented for investigating
fluctuations of a different flavor, i.e. fluctuations in flavor
\cite{qm2002gavin} and baryon number \cite{qm2002wong}.

Charge fluctuations, charge correlations and balance functions represent
another class of observables, which are sensitive to the dynamics of separating
conserved charges. Since charge conservation is local in space-time, each
positive charge is accompanied by a negative charge. If these balancing charges
remain close to one another in coordinate space at the time of their emission,
they will be correlated in momentum space. This means that the net charge
measured into a large volume of phase space, e.g. the STAR acceptance, would
tend more toward zero since each charge would be canceled by a balancing
partner. A delayed hadronization would lead to more highly correlated balancing
charges since the charges would have less time to separate and would also be
produced after the early stage of the collision where there is a high velocity
gradient pulling particles apart. Since most charge is created at
hadronization, a tighter correlation of balancing charges could serve as a
signal of QGP formation \cite{jeonkoch,bassdanpratt}.

\begin{figure}
\centerline{\includegraphics[width=0.48\textwidth]{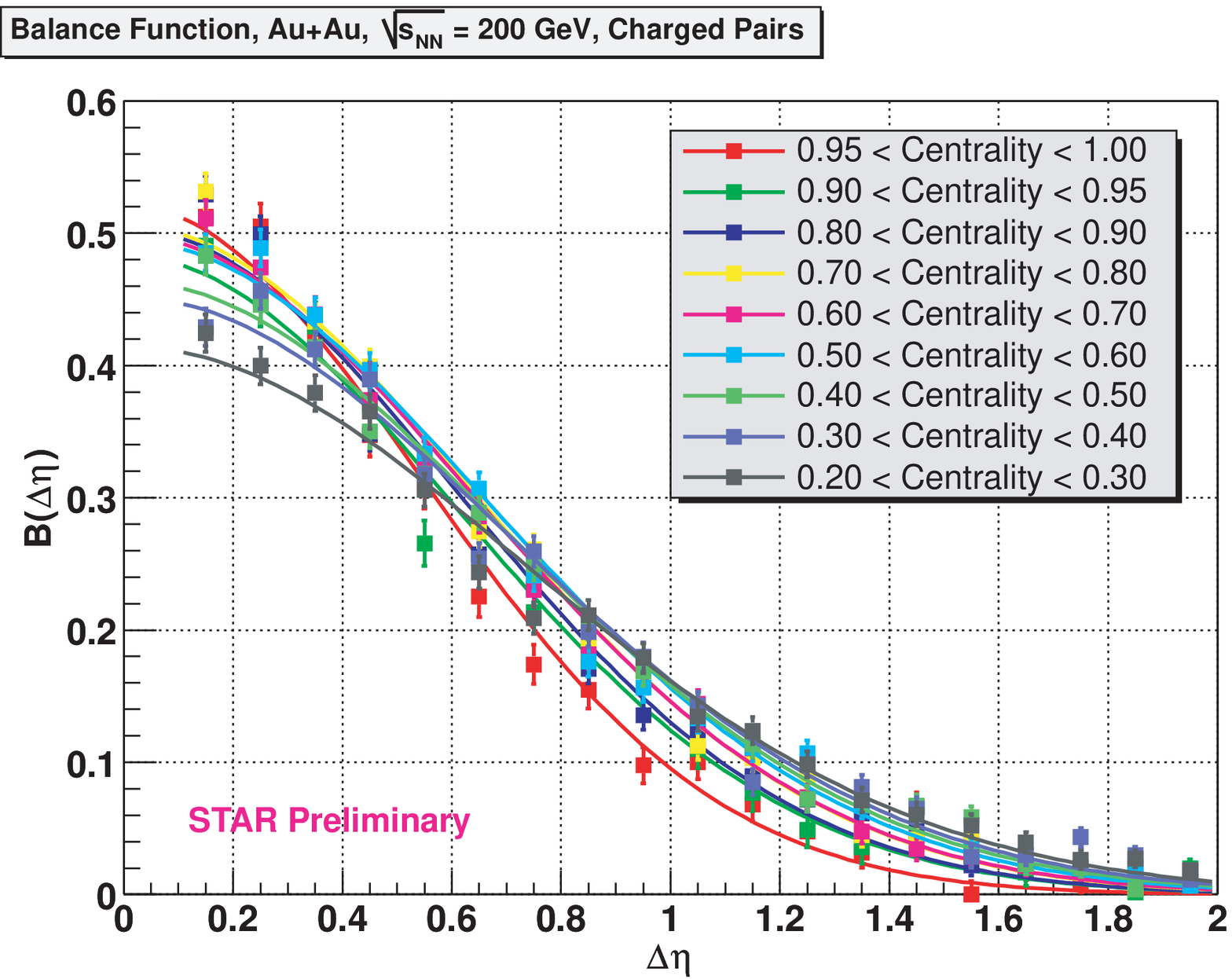}
\hspace*{0.04\textwidth}\includegraphics[width=0.48\textwidth]
{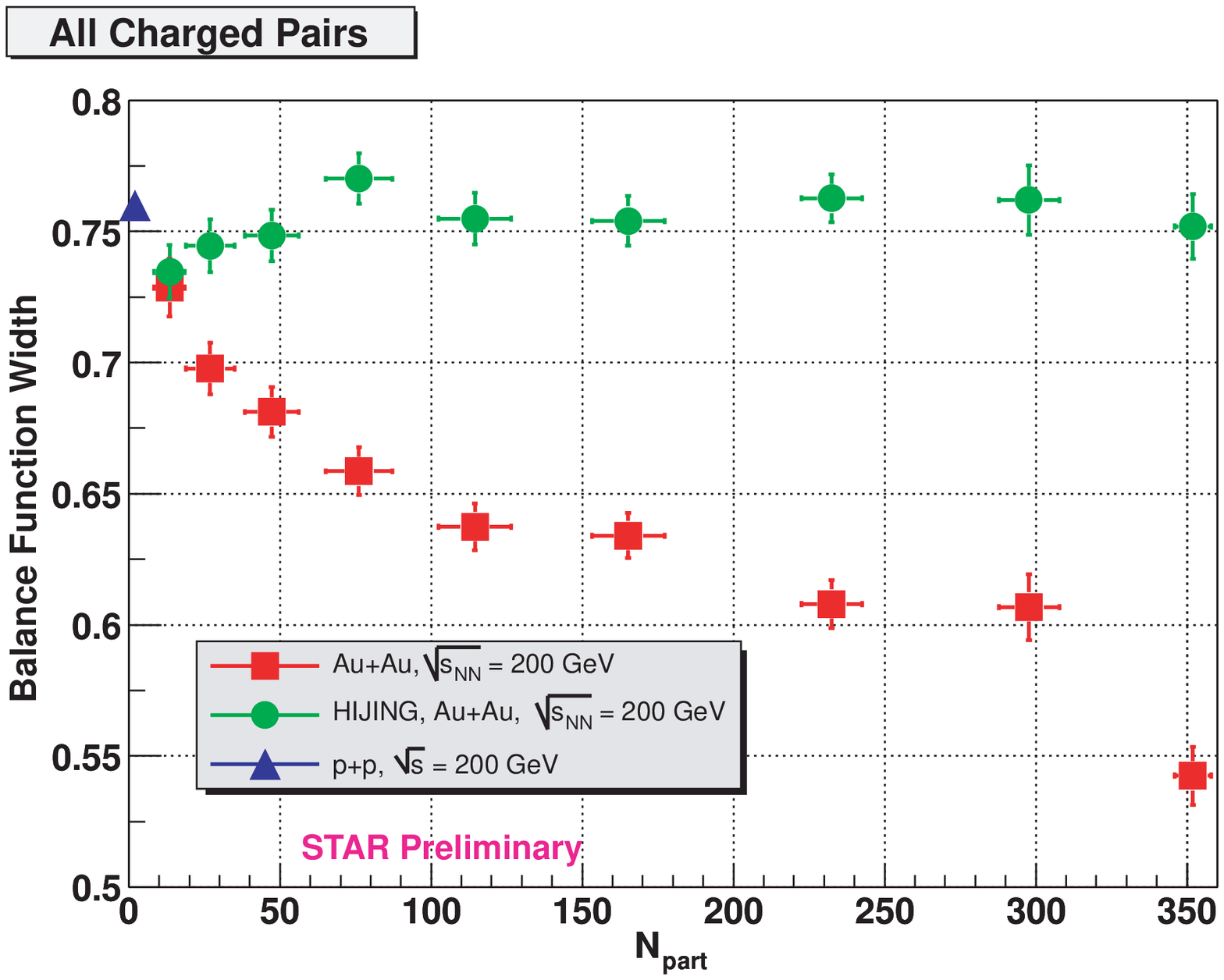}}
\caption{\label{fig:balance}Charge balance functions (left panel) measured by
  STAR show that balancing charges are more correlated in central collisions
  than in peripheral collisions. The widths are plotted as a function of
  centrality in the right panel. This behavior is consistent with a delayed
  hadronization.}
\end{figure}

Several analyses were presented at this meeting which measured this
correlation. Charge fluctuations were measured by NA49 \cite{qm2002blume}, and
charge fluctuations and balance functions were presented by STAR
\cite{qm2002ray}. The balance functions from STAR are shown in
Fig. \ref{fig:balance}. The balance function statistically measures the
distribution of relative rapidities of balancing $\pi^+-\pi^-$ pairs. The
narrowing of the balance function as a function of centrality is highly
suggestive of a delayed hadronization.

\end{document}